\documentclass[aps,prb,longbibliography,a4paper,twocolumn,superscriptaddress,citeautoscript]{revtex4-1}
\usepackage[normalem]{ulem}
\usepackage{hyperref}
\usepackage{color}
\usepackage[english]{babel}
\usepackage{lmodern}
\usepackage[T1]{fontenc}
\usepackage{amsmath}
\usepackage{units}
\usepackage[pdftex]{graphicx}
\usepackage{grffile}
\usepackage{color}
\usepackage[bottom]{footmisc}
\usepackage{graphicx}
\usepackage{dcolumn}
\usepackage{bm}
\usepackage{soul}
\begin{document}
\title{Isotope tuning of the superconducting dome of strontium titanate}
\author{C. W. Rischau, D. Pulmannov\'{a}, G. W. Scheerer, A. Stucky, E. Giannini and D. van der Marel}
\affiliation{Department of Quantum Matter Physics, University of Geneva, 24  Quai Ernest-Ansermet, 1211 Geneva, Switzerland}
\date{\today}
\begin{abstract}
Doped strontium titanate SrTiO$_3$ (STO) is one of the most dilute superconductors known today. The fact that superconductivity occurs at very low carrier concentrations is one of the two reasons that the pairing mechanism is not yet understood, the other is the role played by the proximity to a ferroelectric instability. In undoped STO, ferroelectric order can in fact be stabilized by substituting $^{16}$O with its heavier isotope $^{18}$O. Here we explore the superconducting properties of doped and isotope-substituted SrTi$(^{18}$O$_{y}^{16}$O$_{1-y})_{3-\delta}$
for $0\le y \le 0.81$ and carrier concentrations between $6\times 10^{17}$ and $2\times 10^{20}$ cm$^{-3}$ ($\delta<0.02$).
We show that the superconducting $T_c$ increases when the $^{18}$O concentration is increased. For carrier concentrations around $5\times 10^{19}$~cm$^{-3}$ this $T_c$ increase amounts to almost a factor $3$, with $T_c$ as high as 580~mK for $y=0.74$. When approaching SrTi$^{18}$O$_3$ the maximum $T_c$ occurs at a much smaller carrier densities than for pure SrTi$^{16}$O$_3$. Our observations agree qualitatively with a scenario where superconducting pairing is mediated by fluctuations of the ferroelectric soft mode.
\end{abstract}
\maketitle
\section{Introduction}
Liquid helium doesn't become solid at ambient pressure, even for temperatures close to absolute zero. Solidification is inhibited by the zero-point fluctuations (ZPF) of the He atoms due to their small nuclear mass. A similar effect comes into play in the cubic perovskite SrTiO$_3$ (STO), which would be ferroelectric in the absence of ZPF~\cite{edge2015}. Experimentally, the material is instead observed to be a quantum paraelectric insulator with a large dielectric constant ($\epsilon\sim 10^4$) as the ferroelectric ground state is suppressed by the ZPF of the oxygen atoms~\cite{muller1979}. This is a quantum effect and the ZPF amplitude is controlled by $\hbar^2/m$ where $m$ is the ionic mass. Tuning $\hbar$ is impossible, but substituting more than 33 at.\% of the natural $^{16}$O isotope by the heavier $^{18}$O reduces the ZPF sufficiently to stabilize the ferroelectric phase~\cite{itoh1999,rowley2014}. Alternative methods for tuning ferroelectric order in STO are Ca-substitution~\cite{bednorz1984} and strain~\cite{uwe1976}. 
The advantage of the isotope substitution route is that the stoichiometry remains unaffected and modification of the electronic properties is fully obtained through quantum tuning of the ZPF amplitude. Doping a small density of electrons in STO induces superconductivity with superconducting transition temperature $T_c$ ranging between 50 mK (carrier density 10$^{17}$ cm$^{-3}$) and 400 mK (10$^{20}$ cm$^{-3}$) \cite{schooley1964,koonce1967,lin2013,lin2014}. Isotope substitution is also a key test of the BCS theory of superconductivity. In STO the sign of the isotope effect on $T_c$ is opposite to the BCS prediction and the observed isotope effect is an order of magnitude stronger than the BCS prediction~\cite{stucky2016,tomioka2019}, a state of affairs that has been anticipated by Edge {\em et al.} based on a model where soft ferroelectric fluctuations provide the pairing interaction for superconductivity carriers~\cite{edge2015}. 
Further indications that $T_c$ is enhanced in proximity to ferroelectricity comes from strain experiments~\cite{herrera2018,ahadi2019}.
Due to symmetry selection rules, the coupling to the ferroelectric modes  may be too weak to be compatible with experimental $T_c$ values~\cite{ruhman2016}. However, coupling to pairs of these modes is not ruled out by symmetry considerations and in STO this coupling is in fact quite strong~\cite{ngai1974,marel2019}.
\begin{figure}
\includegraphics[width=0.49\textwidth]{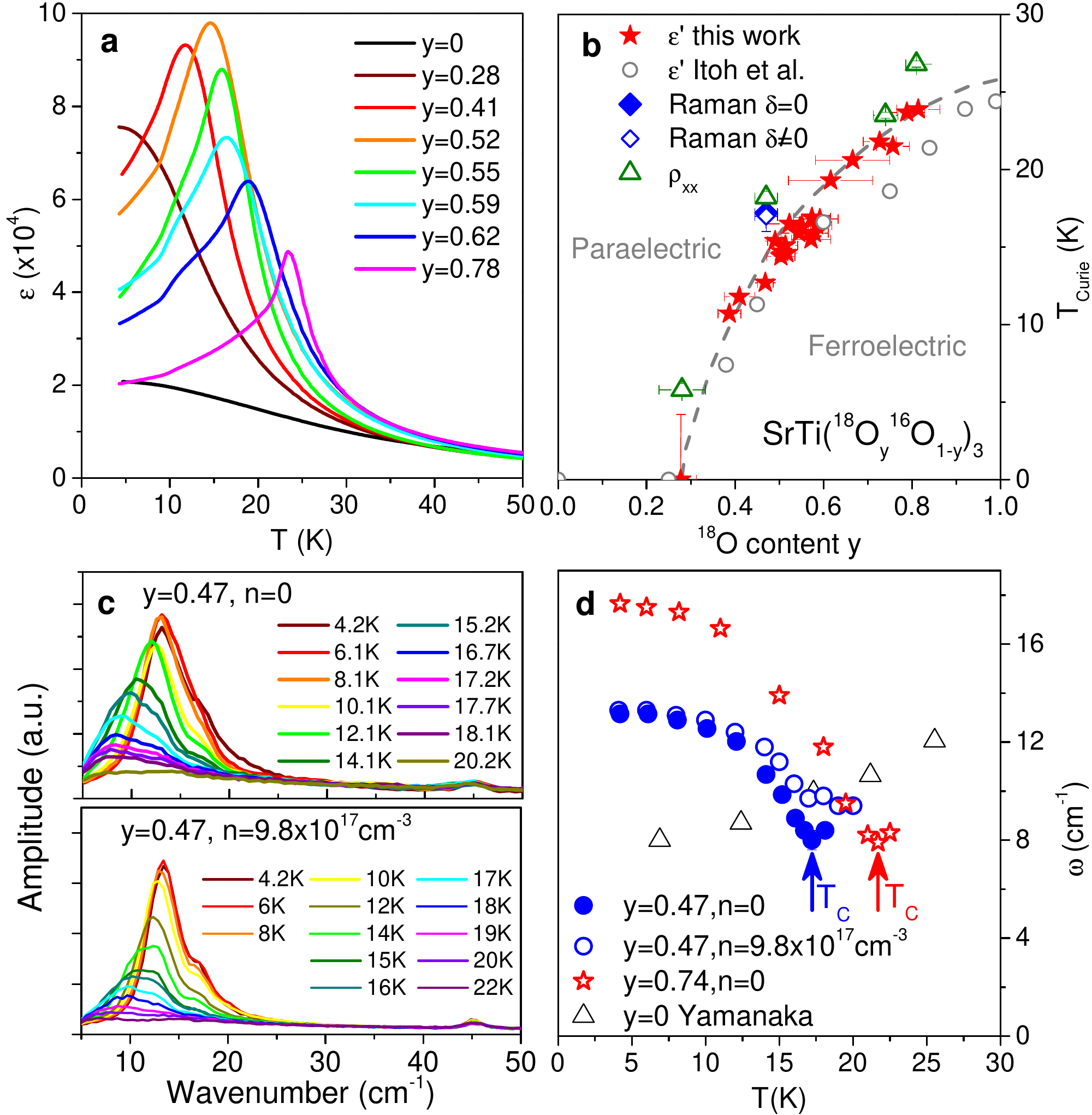}
\caption{\textbf{Ferroelectricity in SrTi$(^{18}$O$_{y}^{16}$O$_{1-y})_{3}$} \textbf{a)} Dielectric constant as a function of temperature. \textbf{b)} Para-ferroelectric phase diagram including previous data by Itoh \textit{et al.} \cite{itoh2003}. \textbf{c)} Low-temperature Raman spectra for $y=0.47$ showing the hardening of the ferroelectric soft mode below the $T_{\textnormal{Curie}}$. \textbf{d)} Temperature-dependence of the Raman-detected soft mode for samples with $y=0.47$ (undoped and doped) and $0.74$ (undoped) compared to hyper-Raman measurements for undoped $y=0$~\cite{yamanaka2000}.
}
\label{Fig1}
\end{figure}
\begin{figure*}
\includegraphics[width=0.99\textwidth]{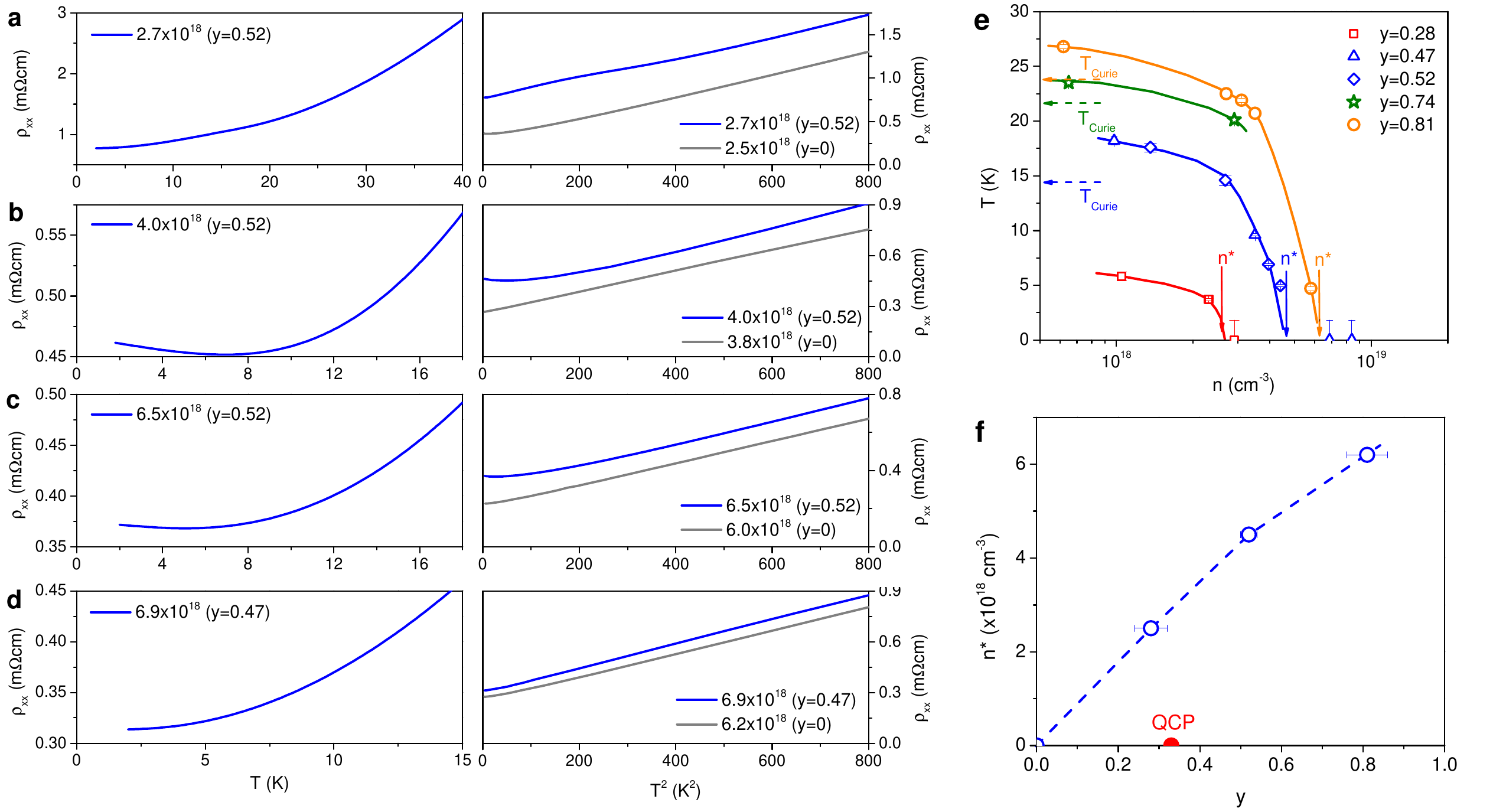}
\caption{\textbf{Temperature dependence of resistivity} \textbf{a)} - \textbf{d)} Resistivity $\rho_{xx}$ as a function of $T$ and $T^2$ for samples with $y\approx0.5$ and different $n$. For comparison, $\rho_{xx}$ of samples with $y=0$ and similar $n$ is shown as well as a function of $T^2$. \textbf{e)} Temperature of the anomaly in resistivity associated with the polar order. Horizontal arrows mark $T_{\textnormal{Curie}}$ measured on the insulating samples with the same $^{18}$O content. \textbf{f)} Critical carrier density $n^{\ast}$ as a function of $y$.}
\label{Fig2}
\end{figure*}
\begin{figure}[!t]
\includegraphics[width=0.49\textwidth]{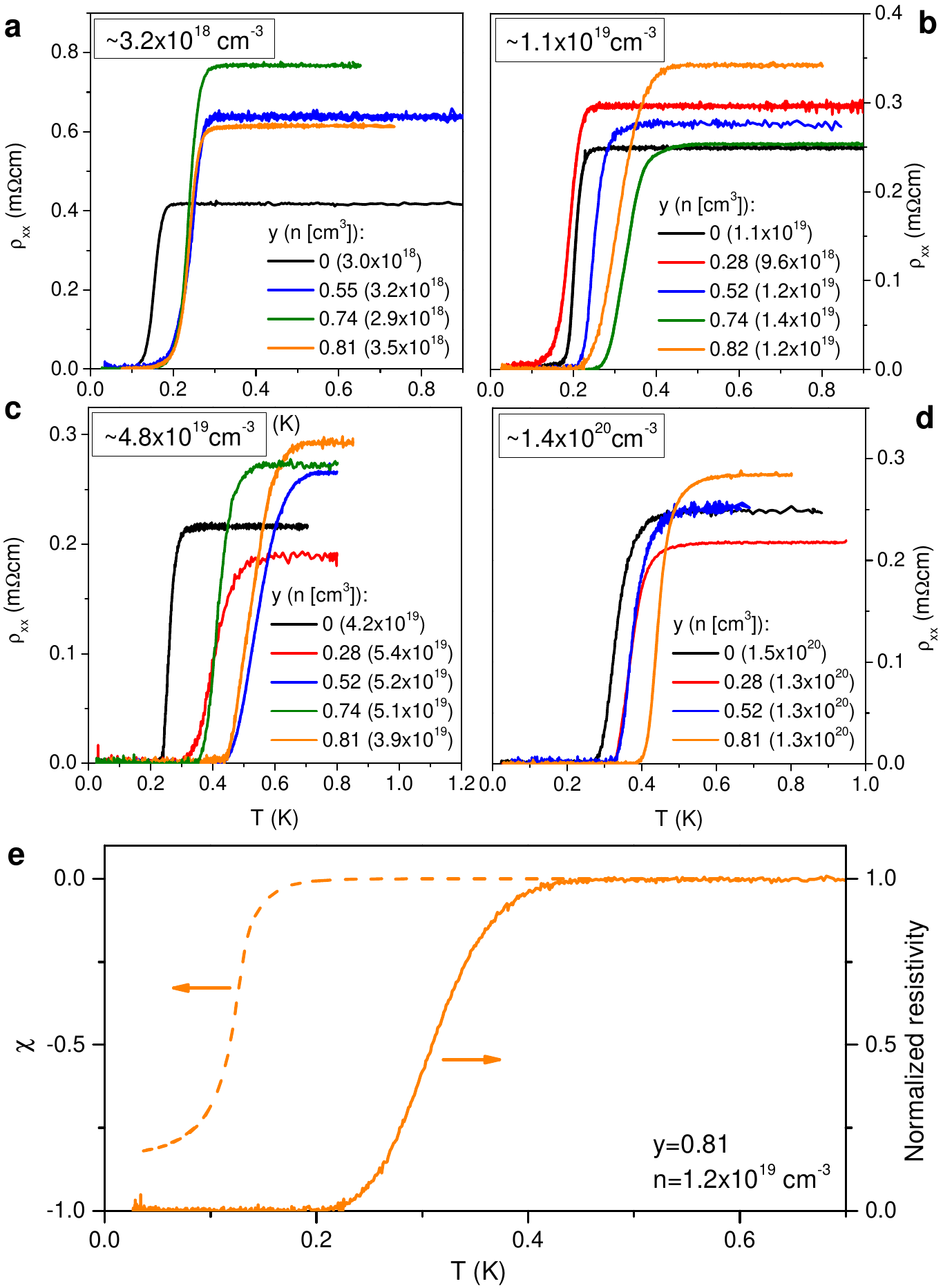}
\caption{\textbf{Superconducting transitions} Superconducting transitions seen in resistivity measurements at selected carrier densities of \textbf{a)} $3.2\times10^{18}$, \textbf{b)} $1.1\times10^{19}$, \textbf{c)} $4.8\times10^{19}$ and \textbf{d)} $1.4\times10^{20}$ cm$^{-3}$ (see Appendix \ref{stransport} for all resistive transitions). \textbf{e)} Superconducting transition seen by a.c. magnetic susceptibility and resistivity for $y=0.81$ and $n=1.2 \times 10^{19}$ cm$^{-3}$.}
\label{Fig3}
\end{figure}
\begin{figure}
\includegraphics[width=0.45\textwidth]{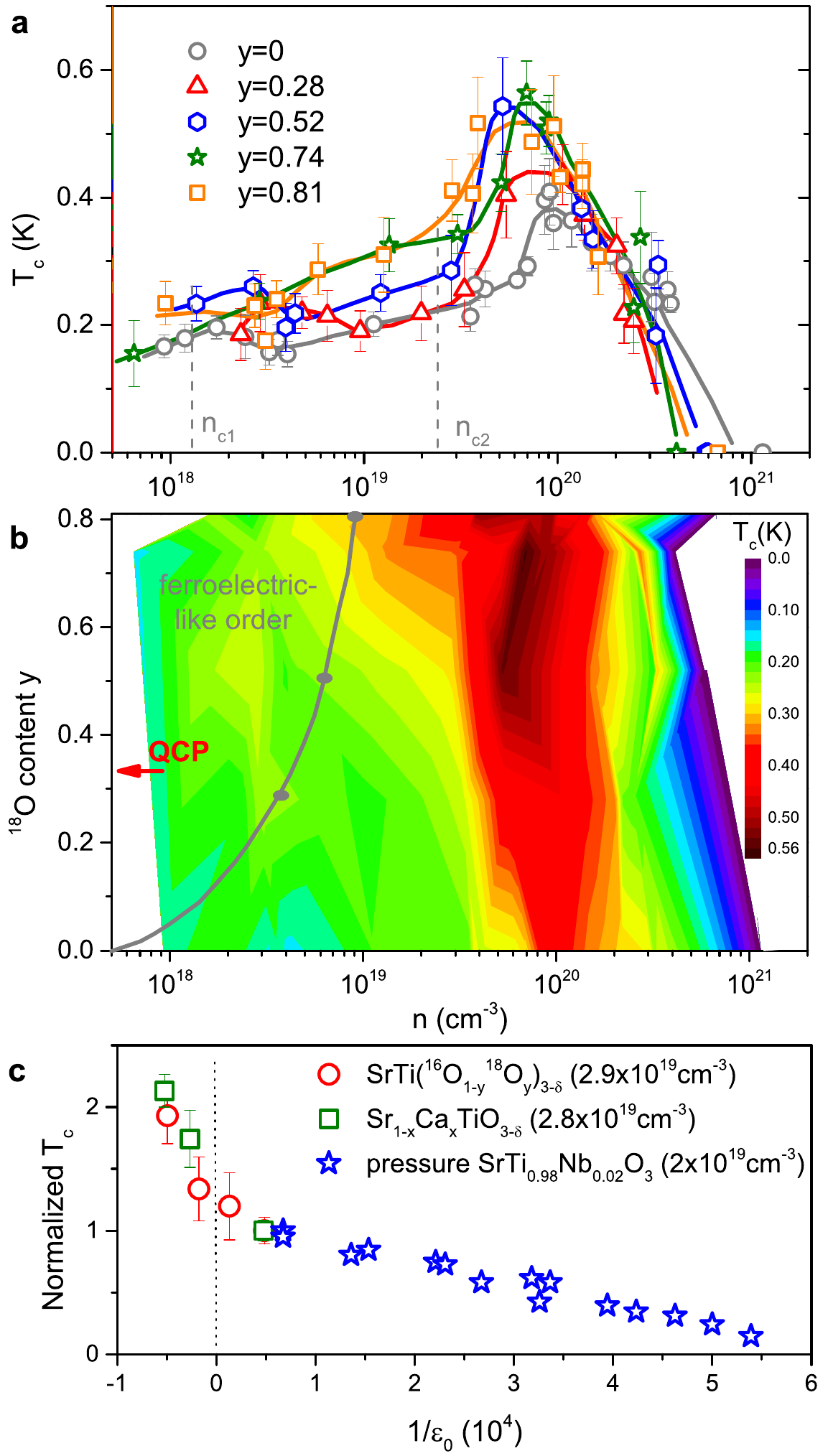}
\caption{\textbf{Superconducting phase diagrams} \textbf{a)} Superconducting $T_c$ as a function of carrier density $n$. Dashed lines mark critical densities $n_{c1}$ and $n_{c2}$ at which the $2^{nd}$ and $3^{rd}$ band are filled, respectively \cite{marel2011,lin2014}. \textbf{b)} Colour plot showing the ferroelectric and the superconducting order as a function of $n$ and $y$. The grey line corresponds to the critical threshold $n^{*}$ (see Fig.~\ref{Fig2}(e)). \textbf{c)} Normalized $T_c$ \textit{vs.} $1/\epsilon_{0}$ taken at low temperature on the insulating samples for oxygen depleted SrTi$(^{18}$O$_{y}^{16}$O$_{1-y})_{3-\delta}$, Sr$_{1-x}$Ca$_{x}$TiO$_{3-\delta}$ (Ref.~\onlinecite{rischau2017} and, in the interest of completeness, an extra measurement for $x=0.0022$)  and SrTi$_{0.98}$Nb$_{0.02}$O$_{3}$ under pressure~\cite{enderlein2020}. Negative $1/\epsilon_{0}$ signifies that insulating samples ($\delta=0$) with same $x$ or $y$ are ferroelectric. $T_c$ was normalized to the value obtained for $y=0$, $x=0$ or zero pressure.}
\label{Fig4}
\end{figure}

Here we provide a broad coverage of the two-dimensional parameter space of  $^{18}$O  isotope substitution (up to 81 \%) and carrier density ($6\times 10^{17}$ - $2\times 10^{21}$ cm$^{-3}$ ) and show how the superconducting dome of STO changes as the system is tuned via $^{18}$O substitution across its quantum critical point (QCP) into the ferroelectric phase (33 at.\% $^{18}$O for undoped STO). Our results challenge the concept of pairing mediated by phonons in the sense of bosonic particles, and require a generalization of the boson concept in the presence of strong anharmonicity.
\section{Ferroelectric order}
We fabricated isotope-substituted SrTi$(^{18}$O$_{y}^{16}$O$_{1-y})_{3}$ samples by heating commercial STO crystals in an $^{18}$O$_2$ atmosphere (see Appendix \ref{Exp} for more details). As-substituted samples are insulating and their dielectric constant as a function of temperature is shown in Fig. \ref{Fig1}(a). Above a certain $^{18}$O level, a peak appears in the dielectric constant marking the Curie temperature $T_{\textnormal{Curie}}$. Fig.~\ref{Fig1}(b) shows the para-ferroelectric phase diagram, i.e., $T_{\textnormal{Curie}}$ as a function of $y$, and confirms previous works that found a ferroelectric order above $y=0.33$ \cite{itoh1999,itoh2003}.

Below 110 K SrTiO$_3$ is tetragonal, and has four infrared active modes. Three modes at 190, 420 and 500 cm$^{-1}$ (TO2, TO3, and TO4 respectively) show less than 6\% redshift for complete $^{18}$O substitution as a result of the heavier mass. The soft ferroelectric mode TO1 at 8 cm$^{-1}$ is by far the most strongly affected due to the strong anharmonicity of this mode.
Raman spectroscopy of the TO1 mode provides additional experimental evidence for ferroelectricity in the isotope-substituted samples.
Fig.~\ref{Fig1}(c) shows Raman spectra for $y=0.47$ displaying the soft mode that becomes Raman active below $T_{\textnormal{Curie}}$ and then shows a hardening with decreasing temperature (see Fig.~\ref{Fig1}(d)) that is not present for $y=0$~\cite{yamanaka2000}. The fact that the ferroelectric soft mode becomes Raman active signals the breaking of inversion symmetry at the Ti sites, which is consistent with ferroelectric order.
\noindent

In a second step isotope-substituted samples with $y=0, 0.28\pm0.04, 0.52\pm0.02, 0.74\pm0.04$ and $0.81\pm0.05$ are annealed in vacuum to remove part of the oxygen atoms to introduce $n$-type carriers and make the system metallic. This diffusive process removes $^{16}$O and $^{18}$O atoms to a different degree due to a (small but finite) difference in the diffusion constant. The effect on the $^{18}$O/$^{16}$O ratio is negligible because only $10^{-5}$ to $10^{-2}$ oxygen atoms/u.c. are removed to make the samples metallic, as compared to $y=0.3$ needed for ferroelectricity. Our metallic SrTi$(^{18}$O$_{y}^{16}$O$_{1-y})_{3-\delta}$ samples display quantum oscillations of the same frequency as previous studies of SrTi$^{16}$O$_{3-\delta}$ 
with the same carrier density~\cite{lin2014,collignon2019} (see Appendix \ref{SdH}). This implies the existence of a well-connected Fermi sea excluding a collection of metallic puddles and shows that the shape of the Fermi surface is not altered upon $^{18}$O substitution. Besides the quantum oscillations, other normal state properties such as low temperature mobility do not change upon isotope substitution (see Appendix \ref{ntransport}).

For the soft ferroelectric mode we observe the same temperature dependence of the Raman spectra in the electron doped material as in the undoped sample (Figs. \ref{Fig1}c and \ref{Fig1}d). This confirms that in these materials ferroelectric inversion symmetry breaking and free carriers can coexist~\cite{rischau2017,wang2019,russell2019}. Since the doped materials are not true ferroelectrics due to the presence of free charge carriers, we will use the label “ferroelectric-like” in those cases.

Below 100~K, the resistivity of $n$-doped STO shows a quadratic temperature dependence of the form $\rho_{xx}=\rho_0 + AT^{2}$. The residual resistivity $\rho_0$ describes the elastic scattering by disorder and the doping-dependent prefactor $A$ the inelasting scattering. 

Figures \ref{Fig2}(a)-(d) show the normal-state resistivity as a function of $T$ and $T^2$ of selected substituted samples. At low temperatures, $\rho_{xx}$ of samples with carrier density below a certain critical threshold $n^{\ast}$ deviates from a $T^2$ behaviour and shows a doping-dependent anomaly that is absent in unsubstituted STO. As can be seen in Fig. \ref{Fig1}(b), the resistivity anomaly occurs around the Curie temperature determined by permittivity and Raman spectroscopy measurements indicating a common cause of resistivity anomaly and Raman activation of the soft FE mode. In Fig.~\ref{Fig2}(e) the temperature of this anomaly (see Appendix \ref{SIRes}) is displayed as a function of carrier density. With increasing $n$, the anomaly shifts to lower temperatures until it disappears at $n^{\ast}$ at which the resistivity recovers a $T^2$ behaviour and the prefactor $A$ is similar for unsubstituted and substituted samples (see Fig.~\ref{Fig2}(d)). Figure \ref{Fig2}(f) shows $n^{\ast}$ as a function of $y$. For $^{18}$O isotope substituted La$_{1-x}$Sr$_x$TiO$_3$ such an anomaly has not been reported~\cite{tomioka2019}, but for Sr$_{1-x}$Ca$_{x}$TiO$_{3-\delta}$  similar features in the temperature dependence of resistivity, thermal expansion and sound velocity, as well as the activation of the TO1 mode in the Raman spectrum, indicate that a ferroelectric-like phase persists in the metallic state \cite{rischau2017,wang2019} and that this phase disappears once a critical threshold of around eight electrons/dipole is reached. The dependence on isotope concentration and carrier density in Fig.~\ref{Fig2}(f) indicates that this symmetry breaking is still present for relatively high carrier concentration $n=2\times10^{18}$~cm$^{-3}$ and low isotope concentration $y=0.28$. The length-scale of this ferroelectric-like order cannot be determined from the present experiments.
\section{Superconductivity}
Figures \ref{Fig3}(a)-(d) show the resistive superconducting transitions at selected carrier densities $n$ for the different isotope levels. There is a clear enhancement of the superconducting transition temperature upon isotope substitution. To rule out any other reason for $T_c$ enhancement other than isotope substitution, twin samples have been annealed in $^{16}$O$_2$ atmosphere under the same conditions as in the $^{18}$O substitution process. No enhancement of $T_c$ was detected in these samples and their superconducting properties agree with previous works \cite{collignon2019}. Fig.~\ref{Fig3}(e) shows the bulk superconducting transition detected in a.c. susceptibility for a sample with $y=0.81$. For strontium titanate the $T_c$ measured by bulk probes (susceptibility, specific heat and thermal conductivity) is always lower than the $T_c$ from resistivity measurements \cite{lin2014b,rischau2017}. The most plausible reason is heterogeneity of the resistive $T_c$ within the samples, possibly associated to the formation of tetragonal domains below 110 K.
The superconducting critical temperatures $T_c$, defined as the mid-point of the resistive transitions, are plotted in the superconducting phase diagram in Fig.~\ref{Fig4}(a). $^{18}$O substitution has a significant influence on both maximum value of $T_c$ as well as the shape of the entire dome. Upon $^{18}$O substitution, the maximum $T_c$ is continuously enhanced with increasing substitution level. Whereas $T_c$ on the overdoped side of the dome remains mostly unchanged, there is a large increase on the underdoped side of the dome {\color{black}($5 \times 10^{18} \lesssim n\lesssim 10^{20}$ cm$^{-3}$)}. This leads not only to a slight shift of the dome maximum to lower doping, but a very strong broadening of the entire dome. This increase of $T_c$ limited to the underdoped side of the dome is clearly visible in the colour plot of $T_c$ in Fig.~\ref{Fig4}(b). The increase already appears at $y=0.28\pm0.04$, i.e., around the quantum critical point confirming earlier results on $y=0.35$ samples \cite{stucky2016}. {\color{black}Below $5\times 10^{18}$ cm$^{-3}$ the effect of isotope substitution diminishes.}

For a given carrier concentration, isotope substitution reduces the energy gap of the polar TO1 mode frequency until the gap vanishes at a critical value. Simultaneously, $T_c$ increases on the underdoped side of the dome.  
Applying pressure has the opposite effect to $^{18}$O substitution as it forces the polar TO1 mode to harden and would therefore correspond to a negative isotope effect in Fig. \ref{Fig4}(b). Pressure experiments have shown that $T_c$ decreases linearly as a function of pressure \cite{pfeiffer1970} and that superconductivity is eventually destroyed at 5.5 kbar for SrTi$_{1.98}$Nb$_{0.02}{}^{16}$O$_3$ ($n=3.4 \times 10^{19}$ cm$^{-3}$) \cite{enderlein2020}. Figure~\ref{Fig4}(c) compares the data obtained for $^{18}$O- and Ca-substituted STO together with the pressure studies as it plots normalized $T_c$ as a function of the inverse dielectric constant $1/\epsilon_0$ taken at low temperature from the undoped systems. 
\section{Conclusions}
An important question is how the isotope substitution affects the superconducting properties, and what the implications of our experiments are for the mechanism of superconductivity in SrTiO$_3$. The trend of increasing $T_c$ upon isotope substitution and decreasing $T_c$ upon applying pressure corresponds qualitatively to the behavior expected when pairing is mediated by the soft ferroelectric mode~\cite{edge2015}. Although the coupling to the LO1 mode at 22 meV is very small ($\lambda=0.003$)~\cite{devreese2010,meevasana2010}, the main features of the phase diagram have been reproduced in a hybrid model of coupling to plasmons and the LO1 mode \cite{takada1980,ruhman2016,klimin2019,enderlein2020}. In this approach the electron-electron interaction is treated in the random phase approximation. Since this approximation has been shown to  strongly overestimate the stability of the superconducting state~\cite{wierzbowska2005}, further analysis of the hybrid phonon-plasmon model is required taking into account exchange and correlation terms. 
The key experimental observations of this work is the isotope-induced shift of the superconducting dome to lower carrier concentrations and the quantum critical line in the doping-isotope plane confirming the theoretical predictions by Edge {\em et al.}~\cite{edge2015}. Increasing the $^{18}$O isotope content from 33\% to 81\% increases the critical carrier concentration from $n=0$ to $n=10^{19}$ cm$^{-3}$. The $T_c$ enhancement in the underdoped region is caused by the confluence of two trends, namely the fact that $T_c$ tends to 0 in the limit of zero carrier density and the enhancement of $T_c$ at the ferroelectric quantum critical point.  
Ruhman and Lee~\cite{ruhman2016} pointed out that the coupling to the TO1 phonon~\cite{edge2015} is very small and vanishes in the limit of zero doping. This argument does not apply to pairs of TO1 phonons \cite{ngai1974,bussmann1993}, which are expected to couple quite strongly to the electrons and may constitute the main pairing mechanism at low doping~\cite{marel2019}. 
\section{Acknowledgements}
This project was supported by the Swiss National Science Foundation through projects 200020-179157 and CRFS-2-199368. The datasets generated and analyzed during the current study are available in Ref.~\onlinecite{yareta} and will be preserved for 10 years. 
\appendix
\begin{figure}[h]
\includegraphics[width=0.49\textwidth]{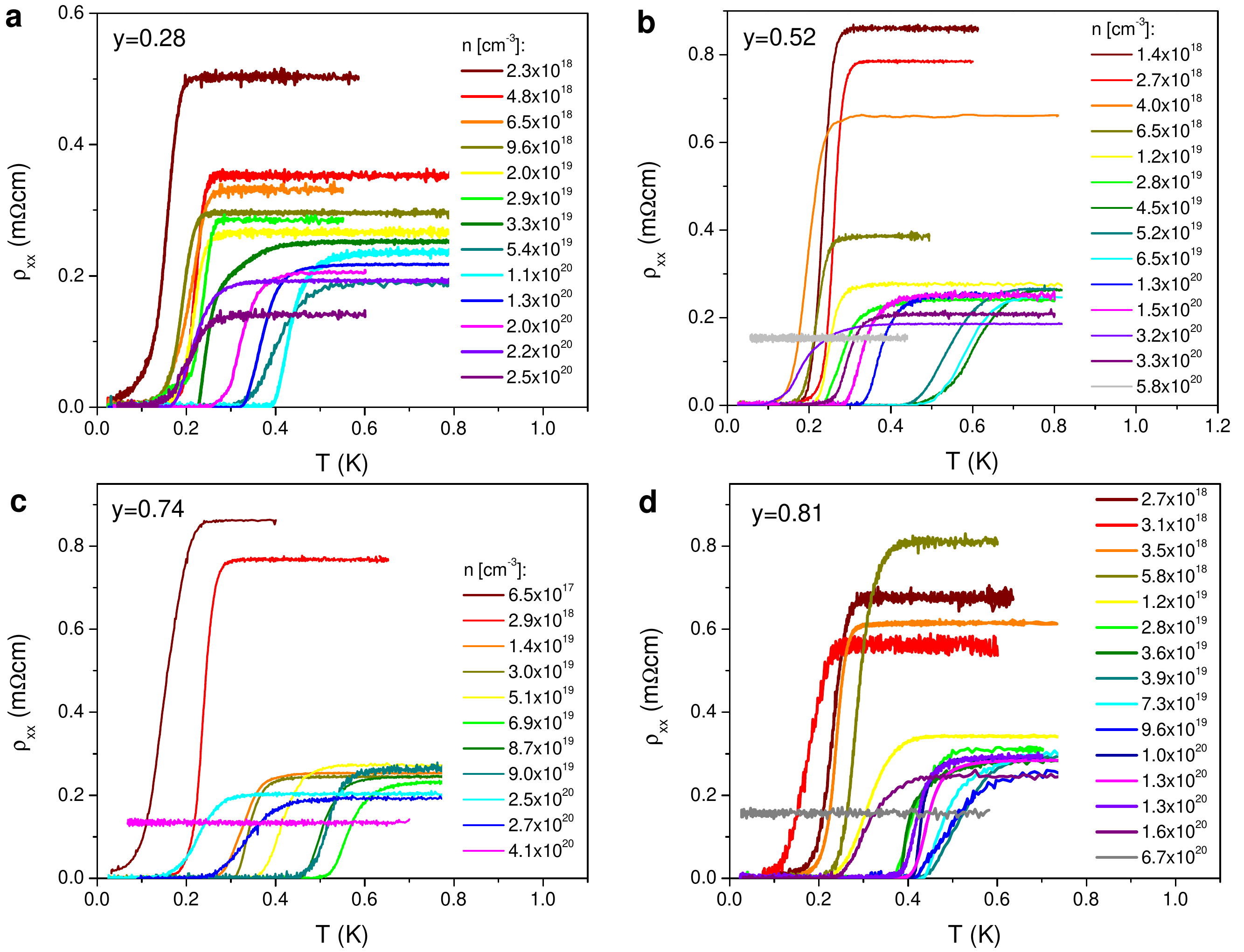}
\caption{\textbf{Resistive superconducting transitions} Resistivity $\rho_{xx}$ as a function of temperature for all measured isotope-substituted SrTi$(^{18}$O$_{y}^{16}$O$_{1-y})_{3-\delta}$ samples.}
\label{figure5}
\end{figure}
\begin{figure}[h]
\includegraphics[width=0.49\textwidth]{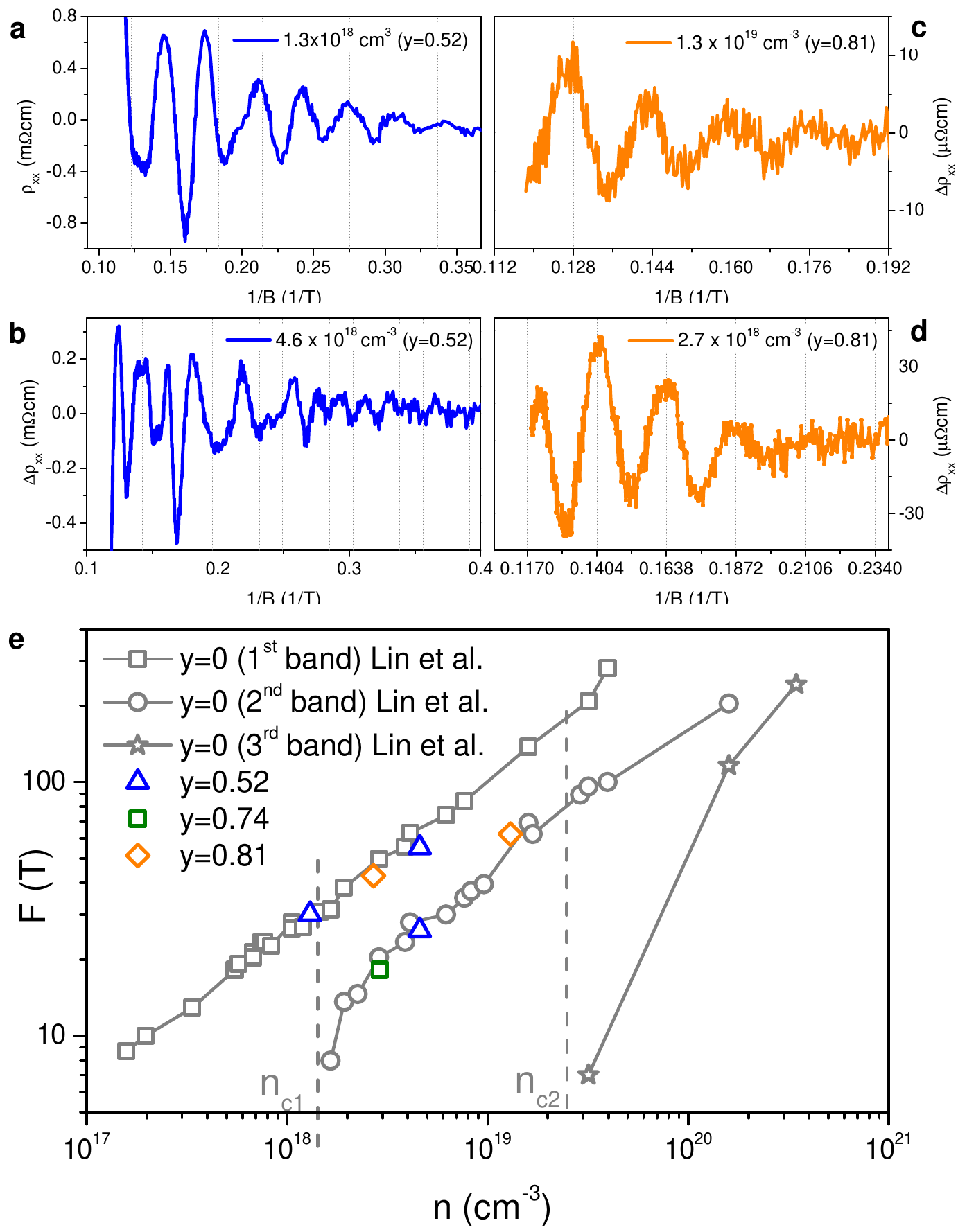}
\caption{Quantum oscillations measured in SrTi$(^{18}$O$_{y}^{16}$O$_{1-y})_{3-\delta}$ samples with $y=0.52$ and $y=0.81$ samples and different carrier densities at a temperature of 30 mK. \textbf{e)} Oscillation frequencies $F$ as a function of carrier density $n$ compared to values measured by Lin et al. \cite{lin2013,collignon2019} on unsubstituted SrTi${}^{16}$O$_{3-\delta}$ $(y=0)$.}
\label{figure6}
\end{figure}
\begin{figure}
\includegraphics[width=0.45\textwidth]{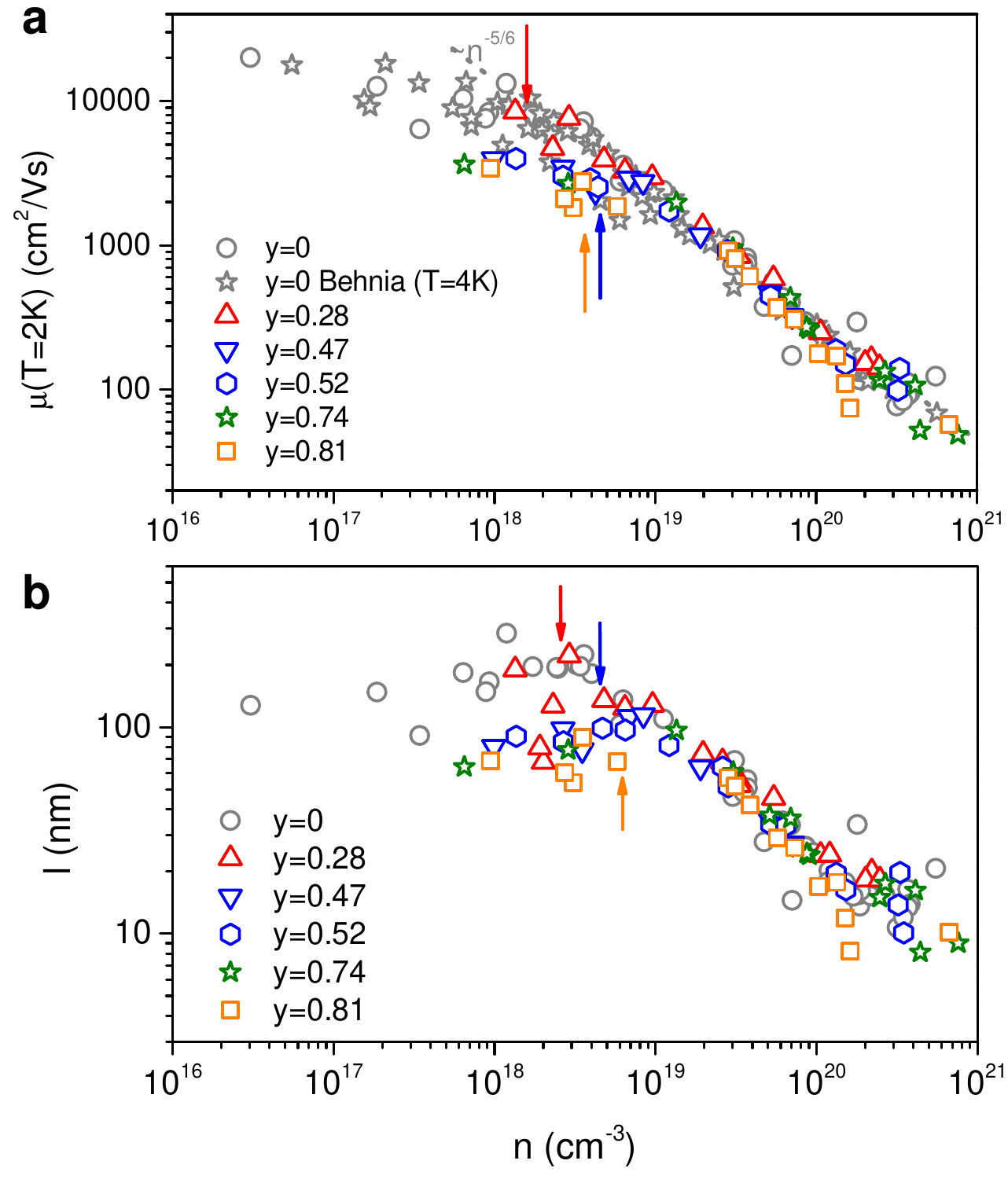}
\caption{\textbf{a)} Low-temperature mobility $\mu$(2K) as a function of carrier density for all samples. The data for $y=0$ is compared to data from ref. \cite{behnia2015} (measured at 4 K). \textbf{b)} Mean-free path $l$ as a function of carrier density. Arrows in both panels mark the critical carrier density $n^{\ast}$ shown in Fig. 2 of the main text.}
\label{figure7}
\end{figure}
\begin{figure}
\includegraphics[width=0.49\textwidth]{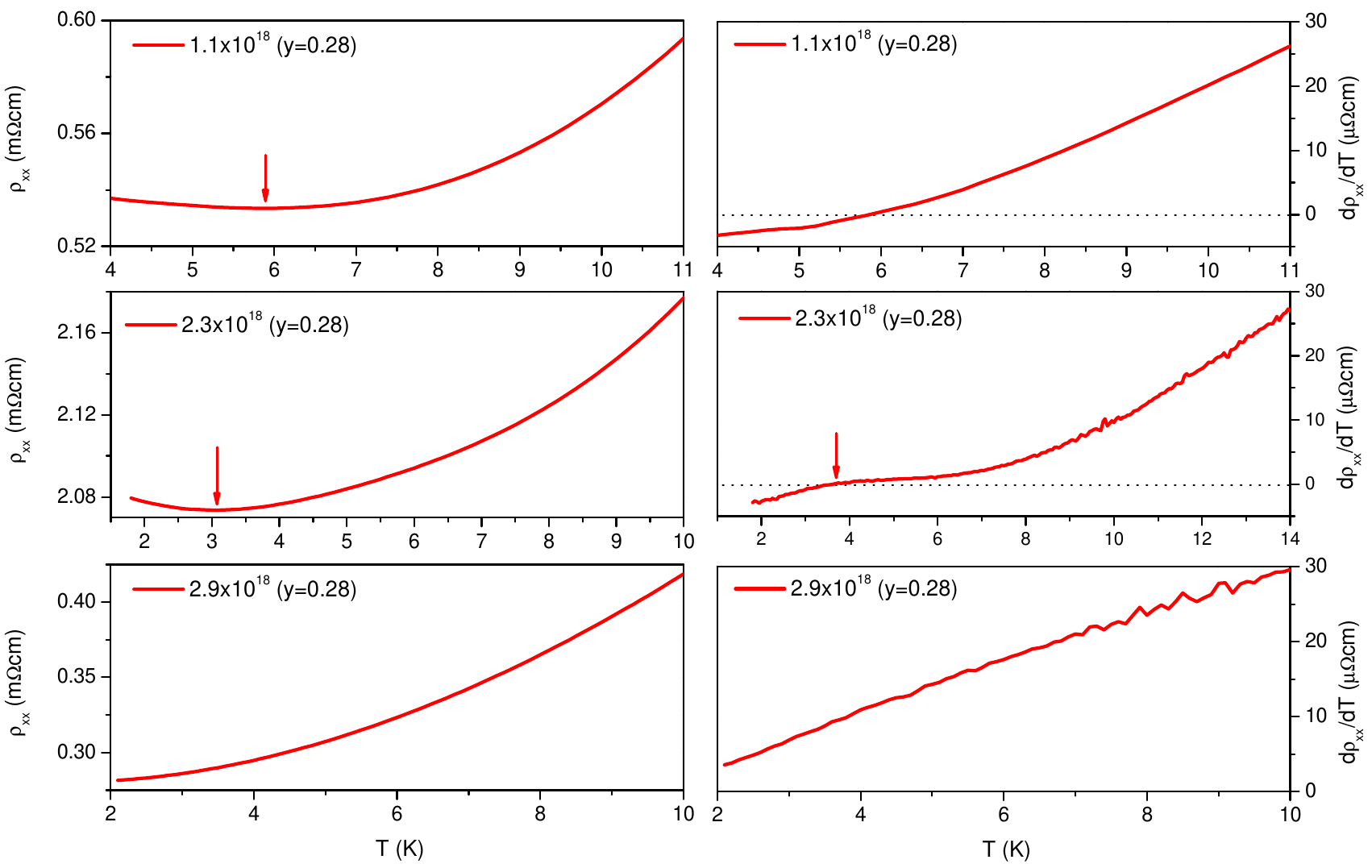}
\caption{Resistivity $\rho_{xx}$ (left hand side) as well as its derivative $d\rho_{xx}/dT$ (right hand side) of samples with an ${}^{18}$O content of $y=0.28$ at different carrier densities $n$.}
\label{figure8}
\end{figure}
\begin{figure}
\includegraphics[width=0.49\textwidth]{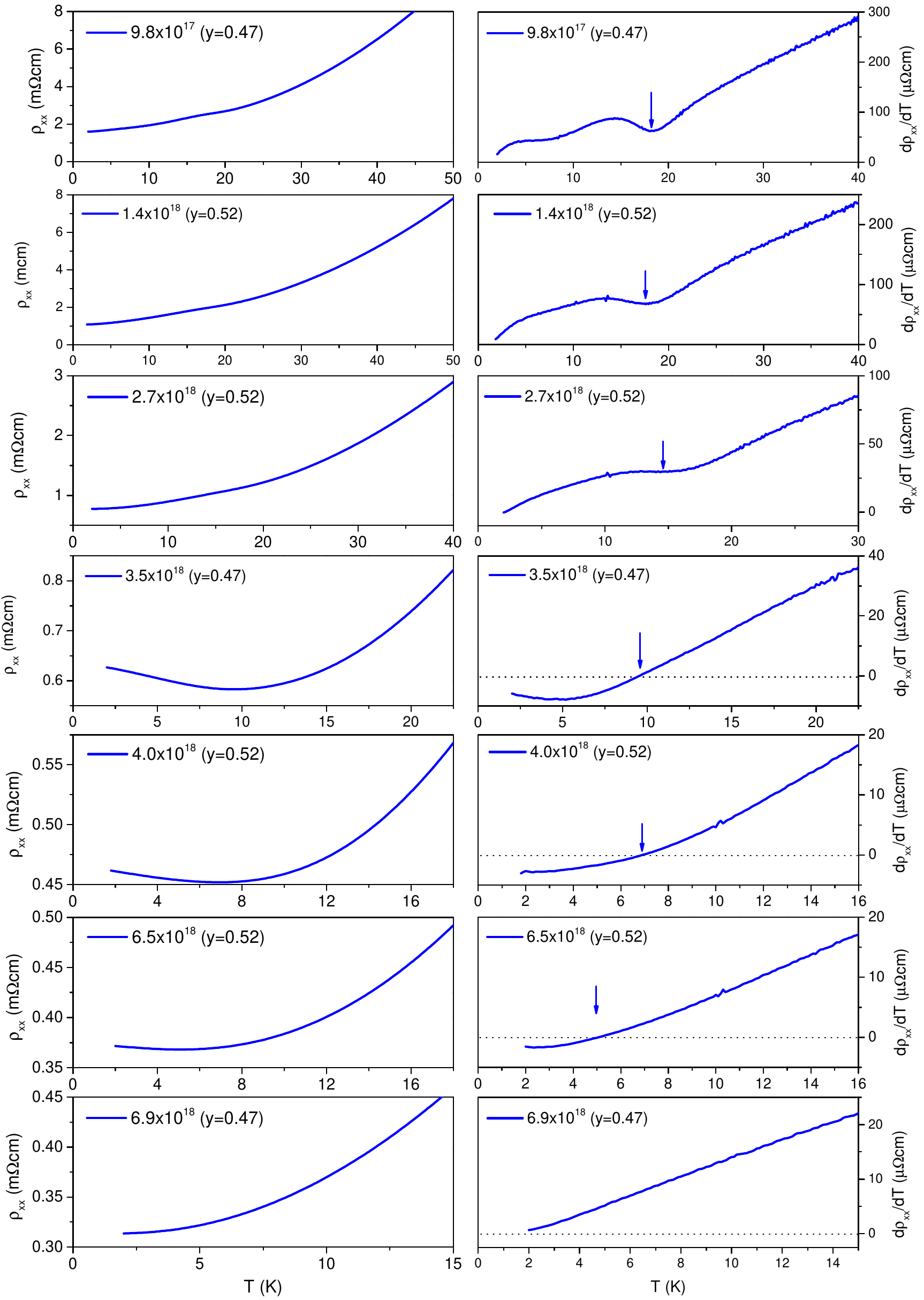}
\caption{Resistivity $\rho_{xx}$ (left hand side) as well as its derivative $d\rho_{xx}/dT$ (right hand side) of samples with an ${}^{18}$O content of $y=0.47$ or $y=0.52$ at different carrier densities $n$.}
\label{figure9}
\end{figure}
\begin{figure}
\includegraphics[width=0.49\textwidth]{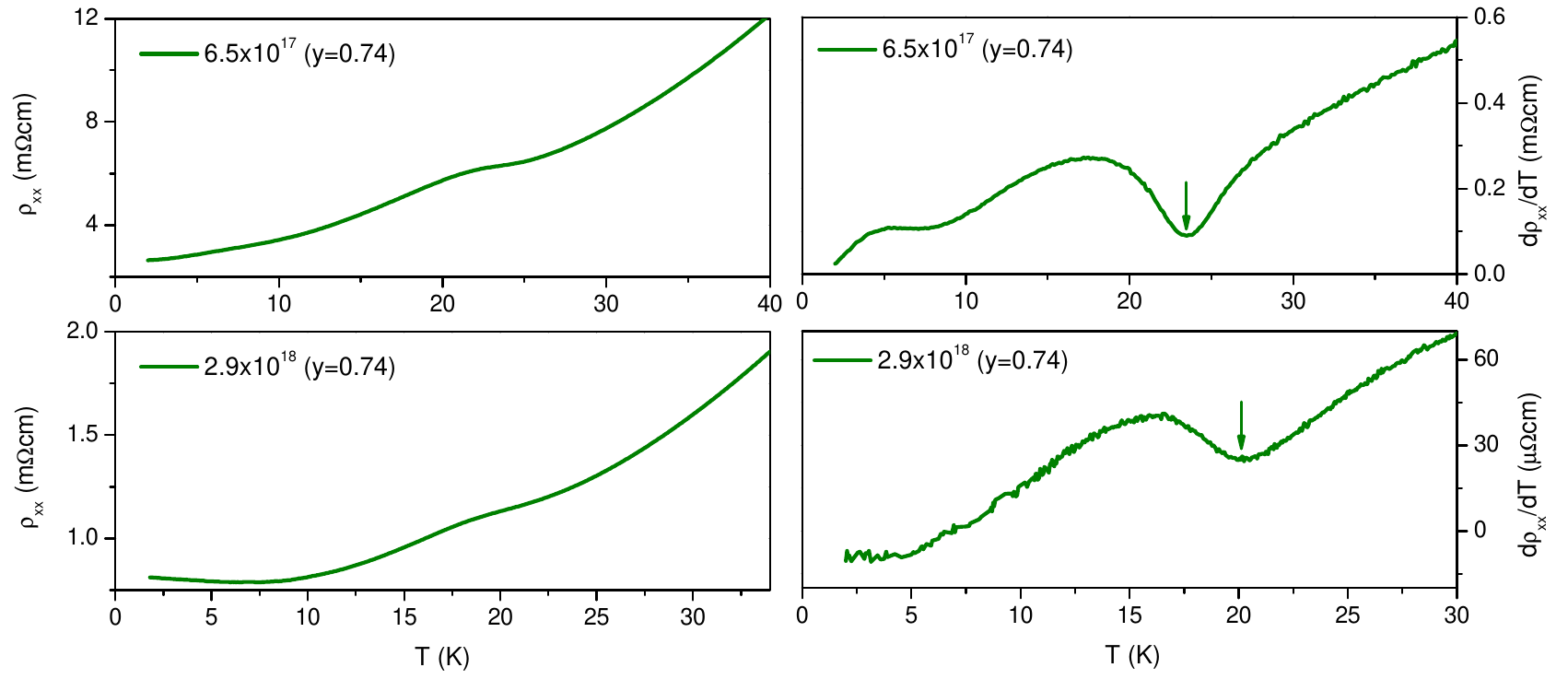}
\caption{Resistivity $\rho_{xx}$ (left hand side) as well as its derivative $d\rho_{xx}/dT$ (right hand side) of samples with an ${}^{18}$O content of $y=0.74$ at different carrier densities $n$.}
\label{figure10}
\end{figure}
\begin{figure}
\includegraphics[width=0.49\textwidth]{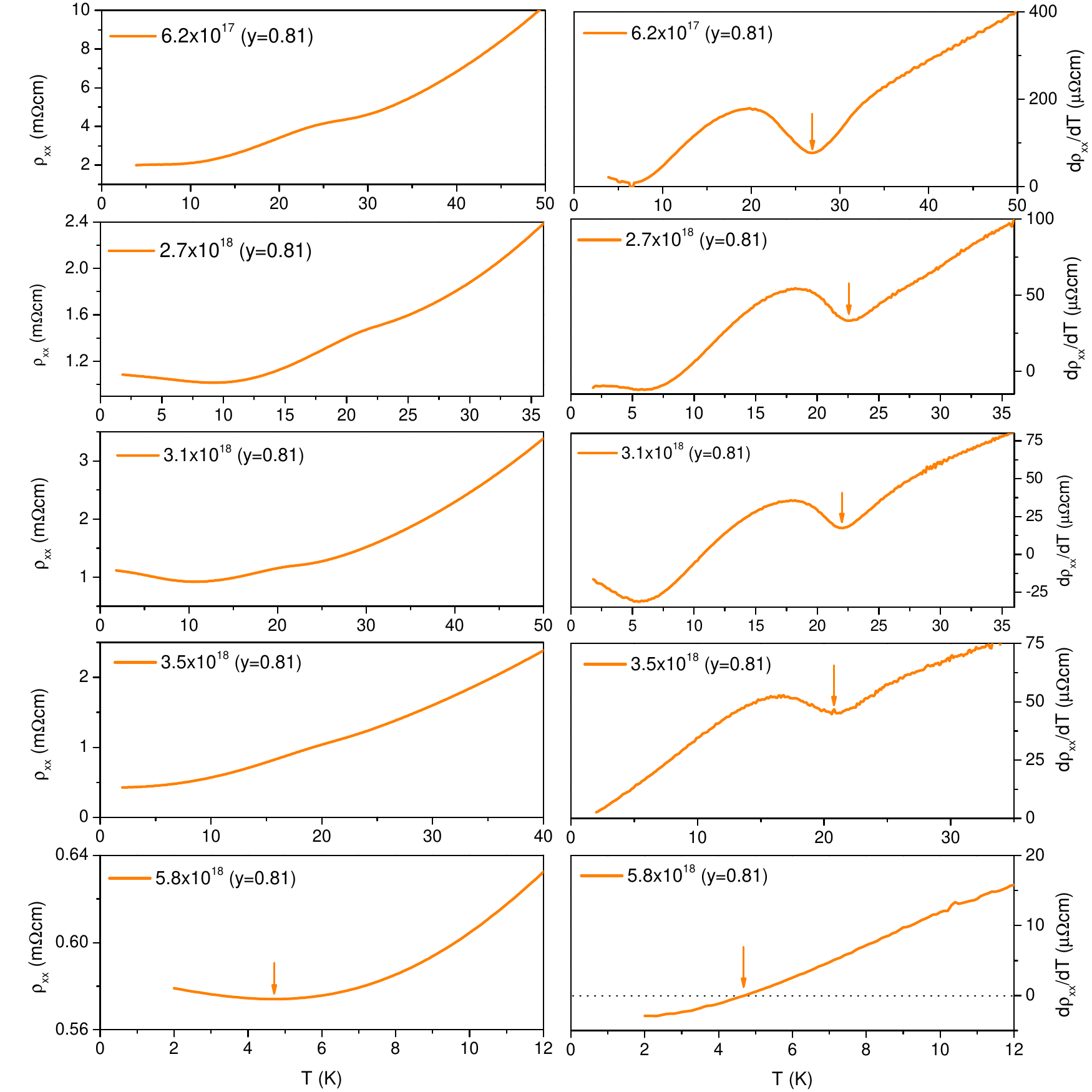}
\caption{Resistivity $\rho_{xx}$ (left hand side) as well as its derivative $d\rho_{xx}/dT$ (right hand side) of samples with an ${}^{18}$O content of $y=0.81$ at different carrier densities $n$.}
\label{figure11}
\end{figure}
\section{\label{Exp}Further information on experimental methods}
In this study we have used commercial SrTi${}^{16}$O$_3$ substrates obtained from different suppliers (MTI, Crystech GmbH and Crystal GmbH). The samples have thicknesses of 0.1 or 0.25 mm and a (100) or (110) crystal orientation. The substitution process has been done in standard quartz tubes that were filled with ${}^{18}$O gas to pressures of 0.4 - 0.7 bar and then sealed with a flame torch. The sealed tubes have been heated using different furnaces at temperatures between 700 and 1100$^\circ$C for at least 20 - 30 days. The annealing time $t$ has been chosen in a way that the diffusion path calculated using $\sqrt{2Dt}$, with $D$ the self-diffusion coefficient of oxygen in strontium titanate \cite{paladino1965,haul1977}, amounts to at least half of the sample thickness. The final ${}^{18}$O concentration in the samples was controlled by the amount of ${}^{18}$O gas in the tube and the number of times the substitution process was repeated.\\ \indent
As in previous studies \cite{itoh1999}, the amount of ${}^{18}$O in the substituted subtrates has been determined by the mass enhancement due to the heavier ${}^{18}$O isotope. The mass of the sample has been measured before and after each substitution process and the substitution rate has been calculated from the mass increase using
\tiny
\begin{equation}
y_f=\frac{m_f[y_iM\textnormal{(SrTi${}^{18}$O$_3$)}+(1-y_i)M\textnormal{(SrTi${}^{16}$O$_3$)}]-m_iM\textnormal{(SrTi${}^{16}$O$_3$)}}{m_i(M\textnormal{(SrTi${}^{18}$O$_3$)}-M\textnormal{(SrTi${}^{16}$O$_3$)})} \nonumber
\label{eq:1}
\end{equation}
\normalsize
with $y_i$ and $y_f$ the initial and final ${}^{18}$O content before and after the substitution process, $m_i$ and $m_f$ the mass of the sample before and after the substitution process and M\textnormal{(SrTi${}^{16}$O$_3$)} and M\textnormal{(SrTi${}^{18}$O$_3$)} are the molar masses of SrTi${}^{16}$O$_3$ and SrTi${}^{18}$O$_3$, respectively. The balance used to measure the weight increase had a precision of $0.01$ mg. The errors of the ${}^{18}$O concentration indicated in the main text stem from the mass variation found in weighing each sample subsequently multiple times. \\ \indent
For the dielectric measurements gold electrodes have been evaporated on both sides of the insulating samples and contacted with silver paste. The capacitance of the samples was measured from 300 to 4.2 K in a home-built cryostat with a high precision LCR meter (Agilent E4980A) using typically an excitation of $50$ mV and 1kHz.\\ \indent
For the electrical measurements (4-point linear resistivity and Hall effect), gold contact pads were evaporated prior to doping by vacuum annealing. Electrical measurements in the temperature range of 1.8 - 300 K have been done in a Quantum Design Physical Property Measurement System (PPMS). The superconductivity of the samples was studied in a dilution refridgerator with a base temperature of 25 mK.
Magnetic a.c. susceptibility measurements were performed in the dilution fridge with a home-built setup using an excitation and a pick-up coil. A lock-in amplifier with a low-noise preamplifier was used to both create the ac excitation current and to read the voltage induced in the pickup coil. The amplitude of the ac excitation fields $B$ was of the order of $1-5$ $\mu$T and had a frequency $f$ in the range of a few kHz. The susceptibility $\chi$ has been calculated from the measured voltage $\Delta V$ drop using $\Delta V=2 \pi^{2} Nf\eta\chi B r^{2}$ with $N$ the number of turns, $r$ the radius of the pickup coil and the filling factor $\eta$ is given by the ratio of the sample volume to the inner volume of the pickup coil \cite{jaccard2010}. It was checked that the measured voltage drop was proportional to both $B$ and $f$ over a wide range of excitation parameters.
\section{\label{stransport}Superconducting resistive transitions}
Figure \ref{figure5} shows the superconducting transitions of all samples displayed in the phase diagram in the main text. The values of $T_c$ shown in the phase diagram correspond to the mid-point of the transition, i.e., the temperature where resistivity is at 50\% of it's normal state value. The error bars of $T_c$ have been defined as half of the temperature difference between the temperatures corresponding to 10 and 90\% of the normal state resistivity.
\section{\label{SdH}Quantum oscillations}
Quantum oscillations in selected samples have been studied at dilution fridge temperatures in magnetoresistance measurements up to 8.5 T. Figures \ref{figure6} a) - d) show the oscillating part $\Delta\rho_{xx}$ of the magnetoresistance, obtained after subtraction of a smooth background, as a function of inverse magnetic field $1/B$. Figure \ref{figure6} e) plots the detected oscillation periods $F$ as a function of carrier density in comparison with frequencies measured by Lin et al. \cite{lin2013,collignon2019} on SrTi${}^{16}$O$_{3-\delta}$. These previous studies of the Fermi surface of doped STO identified two critical doping levels at $n_{c1}\approx1.5 \times 10^{18}$ and $n_{c2}\approx 3\times 10^{19}$ cm$^{-3}$ corresponding to the filling of a second and third band, respectively. Furthermore, they showed that the displayed frequencies match the carrier densities obtained from Hall effect measurements. As shown in Fig. \ref{figure6}, the oscillation periods measured in SrTi$(^{18}$O$_{y}^{16}$O$_{1-y})_{3-\delta}$ samples agree well with those observed for unsubstituted STO, implying that our electron-doped-isotope-substituted samples thus have a single well-connected Fermi surface  with the same shape as observed in unsubstituted SrTi${}^{16}$O$_{3-\delta}$.\\
\section{\label{ntransport}Normal state electric transport}
The low-temperature mobility $\mu={(en\rho_{xx})}^{-1}$ of all investigated samples is shown in Figure \ref{figure7} a). In unsubstituted $n$-doped STO, the mobility is proportional to $\propto n^{-5/6}$ \cite{behnia2015}. In ${}^{18}$O-substituted samples, $\mu$ shows a deviation from this power law behaviour starting below the critical carrier threshold $n^{\ast}$ identified in the main text, i.e., the mobility of the electrons is reduced in the presence of the polar order. Figure \ref{figure7} b) plots the mean free path $l=\hbar \mu k_F/e$  with the Fermi wave vector $k_F=(3\pi^2n)^{1/3}$ obtained by assuming a spherical Fermi surface. Both of the observed features, i.e., a deviation of $\mu$ from a power law dependence below $n^{\ast}$ and a broadened maximum of $l$ around $n^{\ast}$ show that the elastic scattering of the electrons is affected by the polar order that persist below $n^{\ast}$. A similar behaviour of $\mu$ and $l$ has been observed in Sr$_{1-x}$Ca$_{x}$TiO$_{3-\delta}$ samples~\cite{rischau2017,wang2019}. 
\section{\label{SIRes}Temperature dependence of resistivity}
Figures \ref{figure8}, \ref{figure9}, \ref{figure10} and \ref{figure11} show the resistivity $\rho_{xx}$ as well as its derivative $d\rho_{xx}/dT$ for ${}^{18}$O contents $y=0.28$, $0.52$, $0.74$ and $0.81$, respectively, as a function of temperature. {\color{black}Arrows mark the temperature of the anomaly associated with the Curie temperature ferroelectric-like transition at which $\rho_{xx}$ deviates from a $T^2$-behaviour, which is plotted in Fig. 2 of the main text. As the shape of the anomaly depends strongly on the doping level and changes from a hump or kink at lower doping (see for example Fig. 2 a) of the main text) to a minimum followed by an upturn of the resistivity (see for example Fig. 2 b) of the main text), we used the following criteria to define the temperature of the anomaly: For samples that show a clear minimum followed by an upturn, the temperature of the minimum of $\rho_{xx}$ (i.e. the temperature for which $d\rho_{xx}/dT=0$). For samples that show a hump or kink in the resistivity without a minimum, we have taken the temperature of this hump/kink in $d\rho_{xx}/dT$ for this as the hump/kink is more visible in the derivative. Even though other ways of defining the temperature of the anomaly are possible, it does not change the fact that the anomaly disappears at a critical doping threshold $n^{*}$.}

\clearpage
%
%
%
%
\end{document}